\documentclass [12pt]{article}
\usepackage {graphicx}
\usepackage {longtable}

\begin{document}
\begin{center}
\large{Acceleration and ejection of ring vortexes by a convergent
flow as a probable mechanism of arising jet components of AGN}
\end{center}

\bigskip

\begin{center}
\textit{S.A.Poslavsky}$^{1}$\textit{,
E.Yu.Bannikova}$^{1,2}$\textit{, V.M.Kontorovich}$^{1,2}$
\end{center}

\bigskip

\begin{center}
$^{1}$V.N.Karazin Kharkov National University
\end{center}

\begin{center}
$^{2}$Institute of Radio Astronomy NAS of Ukraine
\end{center}

\begin{center}
e-mails: \underline {s.poslavsky@gmail.com}, \underline
{bannikova@astron.kharkov.ua}, \underline {vkont@ri.kharkov.ua}
\end{center}

\bigskip
\begin{abstract}
Exact solutions of two-dimensional hydrodynamics equations for the
symmetric configurations of two and four vortices in the presence
of an arbitrary flow with a singular point are found. The
solutions describe the dynamics of the dipole toroidal vortex in
accretion and wind flows in active galactic nuclei. It is shown
that the toroidal vortices in a converging (accretion) flow, being
compressed along the large radius, are ejected with acceleration
along the axis of symmetry of the nucleus, forming the components
of two-sided jet. The increment of  velocities of the vortices is
determined by the monopole component of the flow only. The dipole
component of the flow determines the asymmetry of ejections in the
case of an asymmetric flow.
\end{abstract}

PACS: 47.32.-y; 47.10.Df; 98.54.Cm

Keywords:
 galaxy -- active, jets; vortices -- ring, plane; flow -- accretion

\newpage

\begin{center}
1. \textbf{Introduction}
\end{center}

\bigskip

A large number of works (see, for example, the monograph~[1]) is
devoted to the origin of jets. In the most of them the decisive
role is played by a strong magnetic field~[2-4], or "external", or
arising from the development of instabilities in the plasma of the
accretion disk. This field serves as a guide for the movement of
particles under the action of electromagnetic, centrifugal and
gravitational forces, allowing them to move against the gravity
and carry away the angular momentum that is necessary for the
effective accretion process, which is responsible for the activity
of the nucleus (see discussion and references in reviews~[5-7]).
At the same time, the very possibility of the existence of strong
magnetic fields in the accretion disks around the black holes is
not entirely clear. In this connection, the models of continuous
flow without magnetic field  are also considered, including those,
in which  the structure  of jets resemble the hydrodynamic
tornadoes [8].

The observations, however, show that at small distances from the
nucleus (parsec scales for active galactic nuclei (AGN))  the
emergence of the individual (including the superluminal)
components of the radio jets are observed~[9]. In the model we
discuss here~[10-12] some ejections are derived from the
kinematics of the interaction of vortices and the exposure of the
magnetic field is not required. An important role in this process
has the flow that can affect the velocity of ejection. Therewith,
greater velocities of the ejected components are attained in a
converging (accretion) flow.

We regard a system of toroidal vortices~[10], surrounding the
central part of AGN, the outermost of which is observed as a
"obscuring torus". The vortex motion arises in the torus due to
twisting by the wind and radiation. Due to the flow symmetry the
movement in the torus possesses the dipole character (Fig. 1
in~[10]). In its simplest form, this movement can be represented
as the motion of two opposite rotating vortex rings in a radial
flow. As is known, the dynamics of a vortex ring can be described
as the movement of a pair of point vortices that arise in the
cross-section of the ring (torus) by the plane of symmetry. In our
case it is a symmetrical system of two or four vortices (or two
vortex pairs) in an arbitrary flow with a point singularity.

\newpage

\begin{center}
2. \textbf{Setting of a problem}
\end{center}

\bigskip

We consider the motion of a system of point vortices on a
background flow caused by the stationary singular point, which is
placed at the origin. Following [13, 14], the stream function can
be represented as the sum of two constituents: its regular part
$\psi _{reg} $ which describes the background flow and the
singular part $\psi _{sing}$, which describes the point vortices.

The complex potential of the flow can be represented as a series

\begin{equation}
\label{eq1}
 w_{reg} = C_{0} \ln z + \frac{{C_{1}} }{{z}} + \frac{{C_{2}} }{{z^{2}}} +
... \qquad (z = x +iy)
\end{equation}

\noindent When the flow is symmetric about the axis \textit{Ox},
all the coefficients \textit{C}$_{k}$ in (1) are real. Expressing
the complex potential through the usual velocity potential
$\varphi$ and the stream function $\psi$ according to $w = \varphi
+ i\psi $ and passing on to the polar coordinates $ z = r \cdot
\exp\,(i\theta) $ we obtain the stream function in the form

\begin{equation}
\label{eq2} \psi _{reg} = C_{0} \theta - \frac{{C_{1}}
}{{r}}\sin\theta - \frac{{C_{2} }}{{r^{2}}}\sin 2\theta - ....
\end{equation}

\noindent The radial and azimuthal velocity components of the
background flow are determined by the conditions

$$ v_{r} = \frac{{1}}{{r}}\frac{{\partial \psi _{reg}} }{{\partial
\theta} }, \quad v_{\theta}  = - \frac{{\partial \psi _{reg}
}}{{\partial r}}~. $$

\noindent Combining the monopole at the origin (source or sink),
dipole, quadrupole, ... with intensity $ C_0, C_1, C_2,  $ ... we
can obtain a given distribution of the radial velocity component
at the circle $\mid z \mid = R $ which describes the considered
flow. The system of vortices and the flow has Hamiltonian form.

\begin{center}
3. \textbf{Symmetric motion of two vortex pairs in the flow with
singularity of the type "source + quadrupole + ..."}
\end{center}

\bigskip

Now we consider the dynamics of a system of two vortex pairs in
the stream flow generated by a fixed singular point  provided the
existence of two axes of symmetry (fig.1). This case can be
interpreted as the motion of a point vortex in the right angle
(which sides are impermeable "walls") at the apex of which is
placed the referred hydrodynamic singularity. Note that the
solution of the problem for a symmetric system of four vortices in
the absence of the background flow was found in the classical
Grobli's work (see [15, 16]). In a purely radial flow it admits a
Hamiltonian formulation [11] and the exact solution of the dynamic
problem [11, 12].

\begin{figure}
\centerline {\resizebox{6cm}{!}{\includegraphics{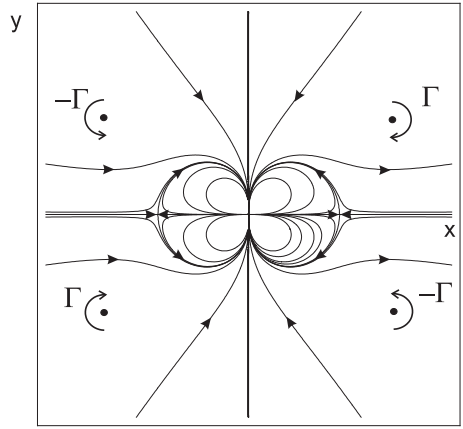}}}
\caption{\textit{\fontsize{12pt}{23pt} Scheme of the movement of
four vortices (two vortex pairs) in a symmetric flow, containing
the sink and the quadrupole for $C_0=-1$,\quad $C_1=0$, \quad
$C_2=1$, \quad $\Gamma=-4 \pi$}} \label{fig1}
\end{figure}

The stream function for the discussed flow is represented in the
form

\begin{equation}
\label{eq3} \psi _{reg} = C_{0} \theta - \frac{{C_{2}}
}{{r^{2}}}\sin2\theta - \frac{{C_{4} }}{{r^{4}}}\sin4\theta - ....
\end{equation}

\noindent The vortex components of the right pair are arranged
symmetrically about the axis \textit{Ox} in the points (\textit{x;
y}) and (\textit{x;-y}), and the intrinsic (due to the interaction
of vortices and not related to the presence of the background
flow) velocity of the vortex in the 1-st quadrant is

\[
\overrightarrow {V} _{sing} = \frac{\Gamma }{4\pi }\left\{
\frac{1}{y} - \frac{y}{x^2 + y^2};\,\,\frac{x}{x^2 + y^2} -
\frac{1}{x} \right\}.
\]

\noindent In the case of only one vortex pair without a background
flow that corresponds to the known expression

\begin{equation}
\label{eq4} \overrightarrow {V} _{s} = \left( {\Gamma /4\pi y;0}
\right),
\end{equation}

\noindent where $\Gamma $ is the intensity of the vortex,
\textit{x} and \textit{y} are  its abscissa and  ordinate.

The dynamics of two pairs of vortices in the flow is described by
the equations

\begin{equation}
\label{eq5} \dot {x} = \frac{\Gamma }{4\pi }\left( \frac{1}{y} -
\frac{y}{x^2 + y^2} \right) + \frac{\partial \psi _{reg}
}{\partial y};\,\,\dot {y} = \left( \frac{x}{x^2 + y^2} -
\frac{1}{x} \right) - \frac{\partial \psi _{reg}} {\partial x}
\end{equation}

\noindent and the Hamiltonian of the system of 4 vortices is
reduced to the form

\begin{equation}
\label{eq6} H = \frac{\Gamma }{4\pi }\ln{\frac{xy}{\sqrt {x^2 +
y^2}}}  + C_0 \arctan\frac{y}{x} - C_2 \frac{2xy}{\left( x^2 + y^2
\right)^2} - ....
\end{equation}

\noindent Accordingly, the equations (5) can be represented as

$$
\dot{x}= \frac{\partial H}{\partial y}, \quad \dot {y} = - \frac
{\partial H}{\partial x}
$$

\noindent Equation ${H = E = const}$  determines the trajectory of
vortices. Moreover, if the motion is unbounded, then the vortex
pair comes from infinity along one axis (\textit{Ox}), exchange
their components and goes away to infinity along the other axis
(\textit{Oy}):
\begin{figure}
\centerline {\resizebox{6cm}{!} {\includegraphics{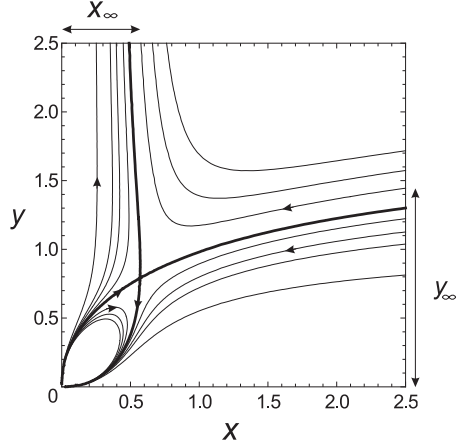}}}
\caption{\textit{\fontsize{12pt}{23pt} Phase portrait for the
vortex of the first quadrant, corresponding to the movement of the
four vortices in the accretion flow. $C_0=-1$, \quad $C_1=0$,
\quad $C_2=1$, \quad $\Gamma=-4\pi$. Thin and bold lines
correspond to the trajectories of the vortex and the separatrix
accordingly. The asymptotic values of coordinates ($x_\infty$ and
$y_\infty$) correspond to half of the distance between the
components of the pairs of vortices at the inlet and outlet of the
system.}} \label{fig2}
\end{figure}
$$
 y_{\infty}=\exp \left\{\frac{4\pi E}{\Gamma}\right\}, \, x \rightarrow +\infty;
 \quad
x_{\infty}=\exp \left\{\frac{4\pi
(E-C_0\frac{\pi}{2})}{\Gamma}\right\}, \, y \rightarrow +\infty;
$$ $$ C_0 \neq 0, \, C_1 \neq 0, \, C_2 \neq 0. $$ For the
asymptotic values $ x_{\infty} $ and $ y_{\infty} $ coordinates of
the vortex moving in the first quadrant, we have the correlation
\begin{equation}
\label{eq7} y_{\infty}  = x_{\infty}  exp\left( {2\pi ^{2}C_{0}
/\Gamma}  \right),
\end{equation}
 which coincides with the result for the purely radial
flow~$^{11,12}$. This means that the ratio $y_{\infty} /x_{\infty}
$ of the limit distances between elements of the vortex pairs at
infinity and, correspondingly, the ratio of their velocities, are
only determined  by the intensity of the $C_{0}$ of the source
(sink) at the origin and do not depend on the other multipole
components.

Since the velocity of the translational motion of a vortex pair is
inversely proportional to the distance between them (cf. (4)),
then we obtain the ratio of the asymptotic values of the
velocities of the pairs, coming from infinity $V_{-}=\Gamma/(4\pi
y_{\infty})$ and going to infinity $V_{+}=\Gamma/(4\pi
x_{\infty})$

\begin{equation}
\label{eq8} V_{ -}  /V_{ +}  = x_{\infty}  /y_{\infty}  =
exp\left( { - 2\pi ^{2}C_{0} /\Gamma}  \right),
\end{equation}

\noindent that coincides with the ratio obtained in~[11, 12].
Obviously, the vortex pairs go away from the source at a slower
velocity, and from the sink -- corresponding to accretion -- with
a greater velocity than they come in. Thus, the obtained solutions
confirm the main result of the work [12] about the acceleration of
ejections by the radial accretion flow, and expand it on the
general case of a symmetric accretion flow. As we will show below
the dipole component of the flow may be responsible for the
asymmetry of ejections.

\begin{center}
4. \textbf{Motion of two vortex pairs in the flow from dipole
which axis is the axis of symmetry of the flow}
\end{center}
\bigskip
In the case of motion of two vortex pairs in the flow from a
dipole with the axis \textit{Oy} with intensity $ C_{1} $ (that
corresponds to replacement  $C_{1} \to iC_{1} $ in (1)) the
equations of dynamics of the vortices can be expressed through the
coordinates of the two vortices in the right half-plane in the
form
\[
\dot {x}_1 = \frac{\partial H}{\partial y_1} \quad \dot {y}_1 = -
\frac{\partial H}{\partial x_1} \quad \dot {x}_2 = -
\frac{\partial H}{\partial y_2 } \quad \dot {y}_2 = \frac{\partial
H}{\partial x_2} ;
\]
\begin{equation}
\label{eq9}
 H = \frac{C_{1} x_{1} }{x_{1} ^{2} + y_{1} ^2} -
 \frac{C_{1} x_{2}}{x_{2} ^{2} + y_{2} ^{2}} +
 \frac{\Gamma} {4\pi}
 \ln\left(
{\left( {x_{1} - x_{2}}  \right)^{2} + \left( {y_{1} - y_{2}}
\right)^{2}} \right) -
\end{equation}
\[
\frac{{\Gamma} }{{4\pi} } \ln\left( {\left( {x_{1} + x_{2}}
\right)^{2} + \left( {y_{1} - y_{2}} \right)^{2}} \right) +
\frac{\Gamma} {4\pi } \ln\,x_1 + \frac{\Gamma} {4\pi} \ln\,x_2 .
\]
\noindent The vortex moving in the first quadrant has the
intensity $\Gamma$ and coordinates $(x_1;  y_1)$ and the vortex in
the fourth quadrant has the intensity -$\Gamma$ and coordinates $
(x_2; y_2) $. The vortices located in the left half-plane have the
parameters -$ \Gamma,(-x_1 ; y_1) $ and $ \Gamma,(-x_2; y_2 )$
accordingly.

From the first integral $ H = E = const $ one can get a
relationship between the asymptotic values $y_{\infty} ,
x_{1\infty} , x_{2\infty} $: the half of the distance
($y_{\infty}$) between the elements of the vortex pairs coming
from infinity along the axis \textit{Ox}, and the same for the
pairs going to infinity along the axis \textit{Oy} ($x_{1\infty} $
and $x_{2\infty} $)

\begin{equation}
\label{eq10} x_{1\infty}  x_{2\infty}  = y_{\infty}  ^{2}.
\end{equation}

\noindent The dipole power occurs implicitly through the
difference of asymptotics \textit{x}$_{1\infty} $ and
\textit{x}$_{2\infty} $. Really, as it is easily seen from the
fig.3, the interaction between the vortices gives the same
distortion in the \textit{x} direction for both  the 1-st and 2-nd
vortices. The difference only arises  due to the influence of the
flow.

\begin{figure}
\begin{center}
 \includegraphics[angle=0,scale=0.65]{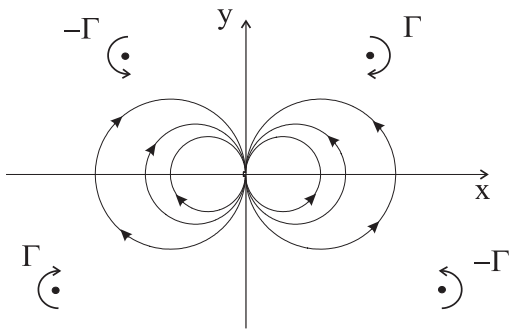}
\end{center}
 \caption{\textit{\fontsize{12pt}{23pt}
 The scheme of motion of four vortices in the dipole flow which
destroys the symmetry relatively of the axis \textit{Ox}. It is
shown the asymmetry of the movement of vortices in such a
stream.}}\label{fig3}
\end{figure}

In the case when there is  also a source (sink)  with intensity
\textit{C}$_{0}$ in the origin, the relationship (10) becomes

\begin{equation}
\label{eq11} x_{1\infty}  x_{2\infty}  = y_{\infty}
^{2}exp\,\left\{ { - \frac{{4\pi ^{2}C_{0}} }{{\Gamma} }}
\right\}.
\end{equation}

\noindent Note that the relation (\ref{eq11}) remains valid if
there are also the multipoles of the higher orders. It has been
taken into account that the initial asymptotics  corresponds to
the symmetric pair of vortices. In terms of the ring vortices this
corresponds to the dipole toroidal vortex of "infinite" radius,
which is compressed by the interaction of the ring components.
Dipole flow introduces an asymmetry in the motion, and different
velocity correspond to vortex rings of different radius thrown in
opposite directions. Thus, within the framework of the
dipole-toroidal model one can naturally  explain both the very
appearance of the ejections accelerated by the accretion flow and
the observed asymmetry of the ejections in AGNs.

\bigskip

\begin{center}
\textbf{Conclusion}
\end{center}

\bigskip

The model of active galactic nuclei proposed in [10]  gives the
possibility within the framework of hydrodynamics of an ideal
incompressible fluid to study the dynamic behavior of toroidal
structures and the influence on their movement of accretion-wind
flows.

In this paper we considered   a simplified plane model, for which
the analogue of a vortex ring is a pair of point vortices, which axis
coincides with the axis of the ring. The dipole-vortex structure
of the torus in the 2D model is represented by the two pairs of
vortices with a common axes and the angular momenta of the
opposite signs. It takes into account that the symmetric pairs of
vortices  correspond to the initial asymptotics.

\begin{figure}
\begin{center}
 \includegraphics[angle=0,scale=0.48]{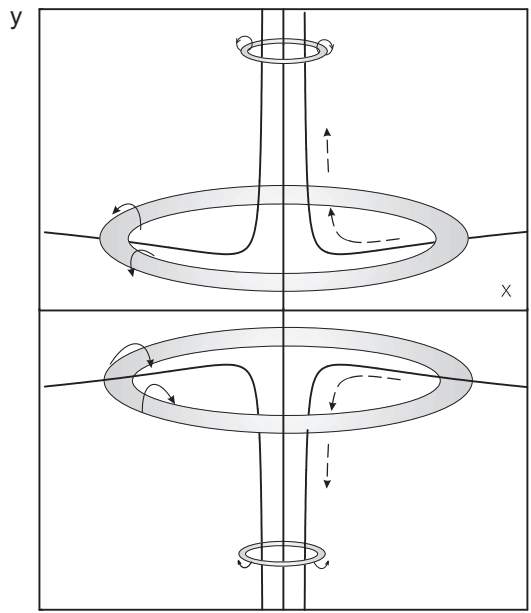}
\end{center}
 \caption{\textit{\fontsize{12pt}{23pt} Scheme of the motion of ring vortices
 in a converging flow,
appropriate to the discussed planar analog (symmetrical flow
[12]).}}\label{fig4}
\end{figure}

In the terms of the ring vortices that corresponds to initial
dipole toroidal vortex of "infinite" radius which is compressed
due to interaction of the ring components. In the absence of the
background flow this problem resembles the classical problem of
the Helmholtz vortex ring interaction with the wall parallel to
the plane in which  the vortex ring lies. The wall can be replaced
by a mirror image of the vortex, and the problem can be reduced to
the interaction of oppositely rotating vortex rings. However, in
our case the direction of rotation is opposite to the
corresponding to approaching the vortex to the wall,  which was
considered by Helmholtz. (Our choice corresponds to the moving off
the vortex from the wall.) In a purely radial flow it admits a
Hamiltonian formulation and exact solution of the dynamic problem
[11, 12]. In this study the conclusion about acceleration of the
ejections by the radial accretion flow is expanded on the general
case of two-dimensional flow. It is shown that  the monopole
component of the flow is only responsible for the acceleration of
ejections, and a dipole flow component may be responsible for the
asymmetry of bilateral ejections.

Thus, the dipole-toroidal model can naturally explain both the
very emergence of ejections, accelerated by the accretion flow in
the case of the general character of the flow, and the observed
asymmetry of the ejections of active galactic nuclei and quasars.

The text basically corresponds to [17]. We have added some figures
and corrected typos in formulas.

For an extended discussion of this subject see [18].

The authors are sincerely grateful to N.N.~Kizilova for the useful
notes.

\newpage

\begin{center}
\textbf{References}
\end{center}

1. V.S.~Beskin. Axisymmetric steady flows in astrophysics, M.:
Fizmatlit, 2006, in Russian; MHD Flows in Compact Astrophysical
Objects, Springer, 2010, 425 pp.

2. G.~Bisnovatyi-Kogan \& A.~Ruzmaikin, Ap \& Space Sci, 42, 401
(1976).

3. R.D.~Blandford, MNRAS, 176, 465 (1976).

4. R.V.E.~Lovelace, Nature, 262, 649 (1976).

5. D.~Lynden-Bell, Mon. Not. R. Astron. Soc. 369, 1167 (2006).

6. R.D.~Blandford, Phil.Trans.R.Soc.Lond. A, 358, 811 (2000).

7. I.F.~Mirabel, Phil.Trans.R.Soc.Lond. A, 358, 841 (2000).

8. M.G.~Abrahamian, Astrofizika, 51, 201, 431, 617 (2008).

9. G.V.~Vermeulen \& M.H.~Cohen, ApJ, 430, 467 (1994).

10. E.Yu.~Bannikova \& V.M.~Kontorovich, Astron. J., 84, 298
(2007), astro-ph/0707.1478.

11. E.Yu.~Bannikova, V.M.~Kontorovich \& G.M.~Resnick, JETP, 132,
¹ 3, 615 (2007).

12. E.Yu.~Bannikova \& V.M.~Kontorovich, Phys.Lett. A, 373, 1856
(2009).

13. G.M.~Reznik, J. Fluid. Mech. 240, 405 (1992).

14. G.~Reznik \& Z.~Kizner, Theor. \& Comp. Fluid Dynamics, 24, \#
1-4, 65-75 (2010).

15. A.V.~Borisov \& I.S.~Mamaev. Mathematical methods for the
dynamics of vortex structures. Moscow-Izhevsk, ICI, 2005, 368 pp.

16. W.~Grobli, Vierteljahrsch. d. Naturforsch.Geselsch., 22, 37,
129 (1887).

17. E.Yu.~Bannikova, V.M.~Kontorovich \& C.A.~Poslavsky, In:
Transformation of waves, coherent structures and turbulence, M:
LENAND, 2009, P.304.

18.  C.A.~Poslavsky, E.Yu.~Bannikova \& V.M.~Kontorovich, 
Astrophysics, Vol. 53, No. 2, 174 (2010)

\end{document}